# The Usefulness of Multilevel Hash Tables with Multiple Hash Functions in Large Databases

**A.T. Akinwale and F.T. Ibharalu**
Department of Computer Science,
University of Agriculture, Abeokuta, Nigeria
atakinwale@yahoo.com

**ABSTRACT:** In this work, attempt is made to select three good hash functions which uniformly distribute hash values that permute their internal states and allow the input bits to generate different output bits. These functions are used in different levels of hash tables that are coded in Java Programming Language and a quite number of data records serve as primary data for testing the performances. The result shows that the two-level hash tables with three different hash functions give a superior performance over one-level hash table with two hash functions or zero-level hash table with one function in term of reducing the conflict keys and quick lookup for a particular element. The result assists to reduce the complexity of join operation in query language from $O(n^2)$ to $O(1)$ by placing larger query result, if any, in multilevel hash tables with multiple hash functions and generate shorter query result.

## Introduction

A hash table is a generalization of an ordinary array which can only accommodate small amount of record data. Record data is growing increasingly due to the large database system. A good query language performance on large data base system requires a quick access to a primary key which may be necessary to select all the needed data records from data base. Similarly, internet user needs a quick access to information. One of the techniques to access these primary keys on data base management system and IP addresses on internet is based on searching. Searching is an operation to find the location of a given item in computer memory. Searching problem is easy to state but nontrivial to solve.

11

There are many searching techniques, for example, *direct chaining* requires a large static data structure, which is not always possible. It has been verified that direct chaining makes effective use of examining an arbitrary position in an array in O(1) time [Aho83]. *Hash function* is employed for reducing the range of array indices that need to be handled. If there are 10,000 data records with **M** slots, then the hash function will handle **M** slots instead of handling data records of 10,000. Each of **M** slots can be in form of linked lists but then searching for an item in these linked lists may take long time.

The aim of this work is to weaken the linked lists so that searching for an item may be accessed at a shorter time. The work will rehash the linked lists into another two hash tables with different hash functions. It will investigate the performance of two-level hash tables with zero-level, and three-level hash tables with two-level. Effort will be made to compare the performance of two-level and one-level hash table against perfect hash functions of the same given **M** slots.

There will be also an effort to generate large numbers of database records in form of randomly generated numbers and unique keys for testing the performance of the hash functions. This will allow us to choose hash functions that will evenly distribute hash values. Ultimately, we will evaluate the performance of multilevel hash tables by using various perfect hash functions.

## 1. Hash Table Functions

Hash Table is an effective data structure which serves to represent dictionary operations. It looks confusing when data is localized in memory in disorder but remains known when it is needed. The idea is to fix hash function $\mathbf{h}: \mathbf{U} \to \mathbf{H}$ where $\mathbf{U} = \{1, ..., \mathbf{n}\}$ on a set of $\mathbf{H} = \{1,...,\mathbf{m}\}$.

The size of **m** determines the size of the hash table, **T**. If the size **m** is short then the size of **T** will also be short. If $\mathbf{m} < \mathbf{n}$ and there exists $\mathbf{k}, \mathbf{i} \in \mathbf{U}$, $\mathbf{k} \neq \mathbf{i}$ and $\mathbf{h}(\mathbf{k}) = \mathbf{h}(\mathbf{i})$, then there is a collision. As far as $\mathbf{m} < \mathbf{n}$, collision is not avoidable. The problem of hash function is collision which has attracted a lot of attentions in research works [Cor92].

## 2. Structure of Multilevel Hash Tables Extension

This research work is devoted to the study of multilevel hash tables which is a specific interesting method of dealing with collisions. The structure of the multilevel hash table extension contains three layers; namely, zero-level hash table with first hash function, one-level hash table with second function and two-level hash table with third function. The first hash function takes a data element as an input and provides an index into an array of buckets usually implemented as linked list of output. If a bucket is allowed to contain another hash table instead of actual keys or (linked list of keys), these keys could be rehashed into another table with a different hash function. This will obviously redistribute the word list (records) instead of placing them into a singly linked list hence there is one-level hash table with a second function.

One-level Hash Table is contained in zero-level hash table, another hash table with linked lists. These linked lists are placed in the buckets and the key contents could also be rehashed into another table with third hash function. This will definitely redistribute the records to form two-level hash tables with third hash function.

Two-level Hash Table contains three-level tables: zero-level hash table, one-level hash table and two-level hash table with the conflict keys in form of linked lists. Figure 1 shows the structure of multilevel hash table extension. As shown in Figure 1, the contents of the bucket at one-level hash table are now in the buckets of two-level hash tables. The contents of buckets at two-level hash tables are now conflict keys in form of linked lists. A key can now be traced along zero-level hash table, one-level hash table and then in the buckets of two-level hash table. This extension of hash tables has definitely weakened the linked lists by reducing the number of conflict keys and searching for record keys now may take a shorter time to locate.

## 3. Selection of Hash Functions for each-level

There are many hashing functions in literature for the purpose of reducing conflict keys and fast computation. Some of these functions perform very well in theory but in practice, their performance is very poor. For example, Universal Hash Function which appears in all algorithms' books of computer science selects hash function at random at run time from a carefully designed class of functions.

$$h_a(x) = \sum_{i=0}^{r} a_i x_i \bmod m$$

This type of hash function may not be appropriate for multilevel hash table extension because it does not uniformly distribute keys above a bucket size of 200. Experiment carried out with a test of data records of 500, 2,000, 5,000 and 10,000 on bucket size of 256 shows that hash values are either distributed at the beginning or in the centre or even at the end of the bucket. Figure 2 shows the result of hash function of 500, 2,000, 5,000 and 10,000 records on 256 buckets. This leaves a lot of memory empty and rehashing of hash values into other tables is empty.

The same experiment is carried out by using hashing codes derived by Robert J. Jenkins' with the same data records and the result shows that hash values are uniformly distributed. Figure 3 depicts Robert J. Jenkins hash codes of 500, 2,000, 5,000 and 10,000 records on 256 buckets. Robert J. Jenkin's hash codes place in every bucket of size 256 the records of 500 as 1 or 2, records of 2,000 as 7 or 8, records of 5,000 as 19 or 20 and records of 10,000 as 39 or 40.

An attempt is made to test other discovered hash functions but none perfectly distributed hash values like Robert J. Jenkin's. For example, Peter J. Weinberger and Arash Partow hash function do not perfectly hash values but uniformly distribute hash values across the buckets more than the size of 256.

Three hashing codes namely, the hashing codes of Robert J. Jenkin, Peter J. Weinberger, and Arash Partow, were selected based on their performance by distributing hash values evenly in buckets of size more than 256.

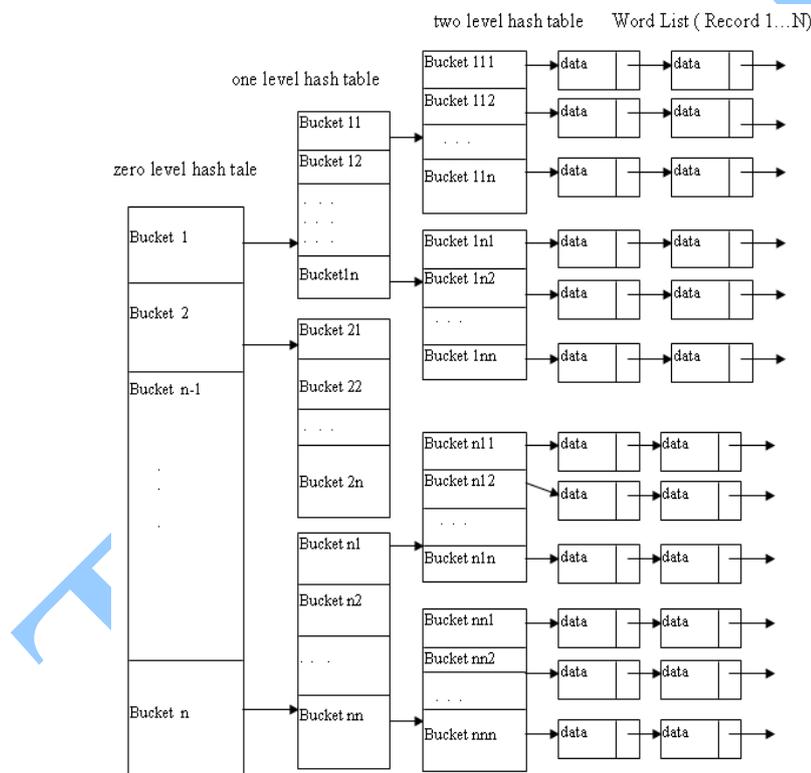

Figure 1: Structure of Multilevel Hash Table Extension

## 4. Instrument of Multilevel Hash Tables Extension

Robert J. Jenkins hash function is used for zero-level hash table; Peter J. Weingeber is employed for one-level hash table while two-level hash table uses

14

Arash Partow hash function. The codes of these three functions are transferred into Java Programming Language.

Test data is of two types: matriculation numbers of students and internet provider addresses. Matriculation numbers of students are made of six alphanumeric data of which two are alphabetic data and the other four are numeric data. For example, a student bearing "strzelecki janek" with identity number 3099 has matriculation number of sj3099. sj is the initial while 3099 is identity number. This number is unique such that there is no student with two matriculation numbers. Similarly, the system generates random number of internet provider addresses in form of $x_1.x_2.x_3.x_4$, where $x_i \in [0...255]$. A test of data records of 500, 2,000, 5,000 and 10,000 were carried out of which matriculation numbers of students and internet provider addresses serve as keys.

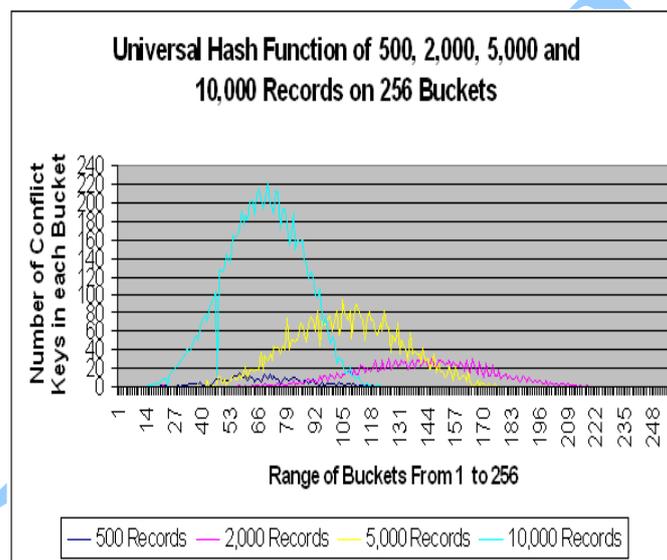

Figure 2: Universal Hash function of 500, 2,000, 5,000 and 10,000 Records on 256 Buckets

## 5. Evaluation of the Results

Robert J. Jenkin's hash function distributes the hash values uniformly into seven buckets. For 500 records, each bucket contains conflict keys of 71 or 72, for 2,000 records each bucket has conflict keys of 285 or 286, record of 5,000 possesses conflict keys of 714 or 715 while 1426 or 1427 conflict keys are in each bucket of 10,000 records.

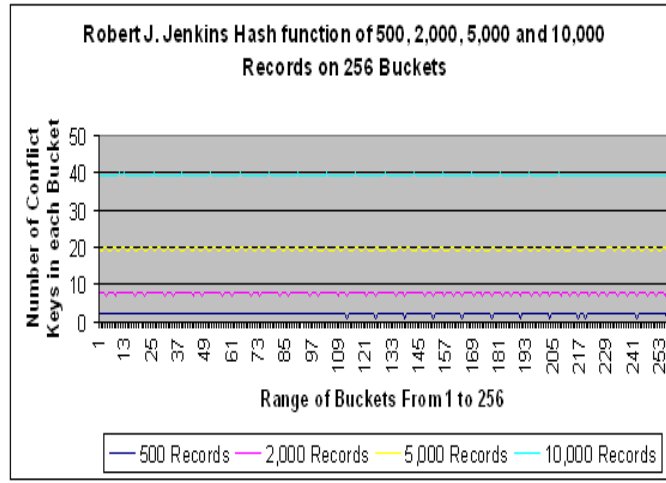

Figure 3: Robert J. Jenkins Hash Function of 500, 2,000, 5,000 and 10,000 Records on 256 Buckets

The hash values are redistributed into one-level hash table using Peter J. Weinberger hash codes. For example, for record 500 at one-level hash table conflict keys of 71 at zero-level hash table are rehashed into 17, 12, 13, 14 and 15. The average case performance is 8 comparisons which is faster than zero-level hash table with 71 comparisons. Similarly, for 2,000 records, 286 conflict keys are redistributed into 51, 66, 51, 64 and 54. For 10,000 records at one-level, conflict keys of 1427 of zero-level hash table are rehashed into 320, 258, 283, 265 and 301. In this step, the conflict keys have been reduced in size. The average case performance is 134 comparisons which is faster than zero-level with 714 comparisons. Table 1 illustrates the summary of average and worst case comparisons of 500, 2,000, 5,000 and 10,000 records using bucket size of 7, 5 and 3. Looking at the table 1, for 500 records at two-level hash table, conflict keys of 12 have been rehashed into 5, 3 and 4. The average case performance is 2 comparisons. For 2,000 records at two-level hash tables, conflict keys of 53 have been redistributed into 15, 22 and 16. It is now easy to traverse the linked lists of 22 at a short time. For 10,000 records at two-level hash table, conflict keys of 320 at two-level hash table have been rehashed into 113, 101 and 106. The average case performance is 39 which is faster than one-level hash table with 134 and zero-level hash table of 714 comparisons. In term of comparison of keys during the traversing the linked lists, the result indicates that one-level is better than zero-level while two-level is far better than either one-level or zero-level.

| level | Size = 500 average | Size = 500 worst | Size = 2,000 average | Size = 2,000 worst | Size = 5,000 average | Size = 5,000 worst | Size = 10,000 average | Size = 10,000 worst |
|---|---|---|---|---|---|---|---|---|
| zero | 36 | 72 | 143 | 286 | 353 | 715 | 714 | 1428 |
| one | 5 | 21 | 22 | 73 | 63 | 171 | 134 | 320 |
| two | 2 | 2 | 5 | 28 | 16 | 67 | 39 | 116 |

Table 1: Summary of average and maximum worst case of the multilevel hash tables.

## 6. Perfect Hash Function versus Multilevel Hash Tables

Robert J Jenkins hash function uniformly distributes keys in various buckets. The question is why do we need to rehash the key?

For 10,000 data records of ( 7 * 5 ) = 35 buckets for perfect hash function, Robert J. Jenkin's hash function places 286 or 287 hash values into each 35 buckets. The average case is 143 comparisons while worst case is 287 comparisons. For one-level hash table with the same parameters, the average case is 134 while worst case is 320 comparisons. Average case of one-level hash table (134) is faster than average case of perfect hash function (143) but the opposite is the worst case. The same analysis is to the perfect hash function of 10,000 data records on (7 * 5 * 3) = 105 buckets, the average case is 48 while 96 is the worst case. Two-level hash table with the same parameters has average case of 39 and 116 for the worst case. Like previous result, only average case of two-level hash table (39) is faster than the average case (48) of perfect hash function. Clearly, as stated by Cormen [Cor92], hash table is not measured by its worst case performance.

## 7. Usefulness of the Results

The usefulness of multiple hash tables with various hash functions can find prominence in database applications where we need to perform primary keys lookup in large data base, and lookup for IP addresses on network.

### 7.1. Lookup for Primary keys

One of the usefulness of multilevel hash table extension with three hash functions is for quick lookup of keys in large data base management system. Join operation in Structural Query language combines two separate databases by comparing each key with all the keys in the second database. The runtime complexity is (**n: m**) = O ($n^2$). Using multilevel hash table extension, a larger database, if any, may be placed on a multilevel hash table with various hash functions which will weaken the conflict keys. Assuming **n** = 2,000 and **m** = 10,000, figure 6 shows join operation in multilevel hash table extension that compares **n** keys with conflict keys m = {113, 101, 106 ...}at two-level hash table which is very short in size. The complexity is O (1). The time to search the conflict keys on the linked list is very short. It is very fast and efficiency. The only problem is memory which linked lists will need to occupy and this can be addressed by adding more memory to the data base management system.

### 7.2. Lookup for IP Addresses

The technique employed on network is to access the address of IP at quick time. Due to the increase in use of network and the desire of the users to access information at short time, attempt is made to make sure the system can access a given IP address very fast. A random number is generated in form of IP addresses from 0:0:0:0 to 125:125:125:125. These numbers serve as keys to the hash functions at-level zero, one and two hash table. Figure 4 and 5 illustrate the time to lookup for a given key. For example, looking at figure 4 with 2,000 records, it takes**13203** milliseconds to lookup for a particular key at zero-level, **7250** milliseconds at one-level while **6219** milliseconds at two-level for the same key. In figure 5 with 10,000 data records, it takes **40328** milliseconds to search for a given key at zero-level, **13203** milliseconds at one-level and **8391** at two-levels. The same analysis is also applied to data records of 500 and 5,000. Results indicate that it is very fast to lookup for a given key at one-level hash table than zero-level while it is faster to search for a particular key at two-level than one-level. This indicates that multilevel hash tables with multiple hash functions can be employed to lookup for IP address in internet environment.

### Conclusion

Three selected hash functions are used for multiple hash tables after testing their performance and ensure that they distribute hash values evenly in above 256 buckets. Each hash function is assigned to different-level of hash tables. The hash functions are written in Java Programming Language with time to lookup for a given particular key. Data records of different sizes are employed to test the performance of the system.

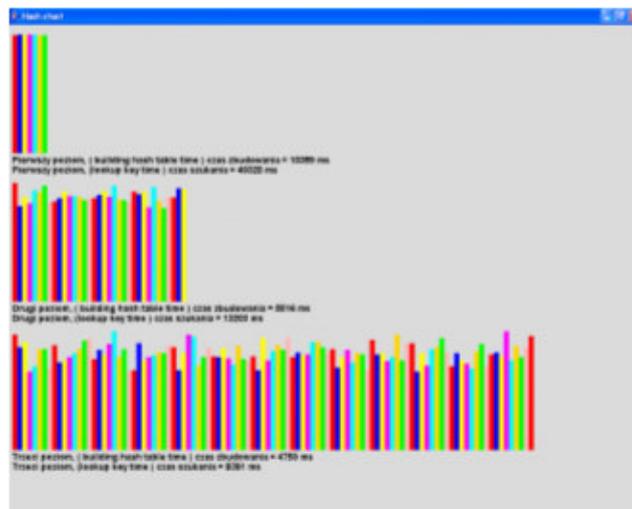

Figure 4: Lookup for a Key in 2000 Data Records

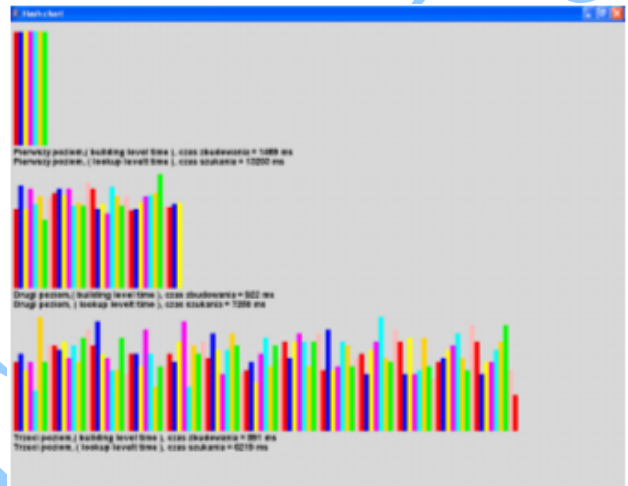

Figure 5: Lookup for a Key in 10,000 Data Records

The selected three hash functions build the input blocks into the respective zero, one and two-level hash tables and provide lookup for a given keys. The results indicate that searching an element at two-level with third hash function perform very well than searching a given item either at one or zero-level. The result also shows that multiple hash tables with multiple hash functions can be employed in join operation of structural query language, by placing the larger query result into multilevel hash tables with multiple hash functions. The result indicates that the system can be used to search for a particular IP address at a quick time.

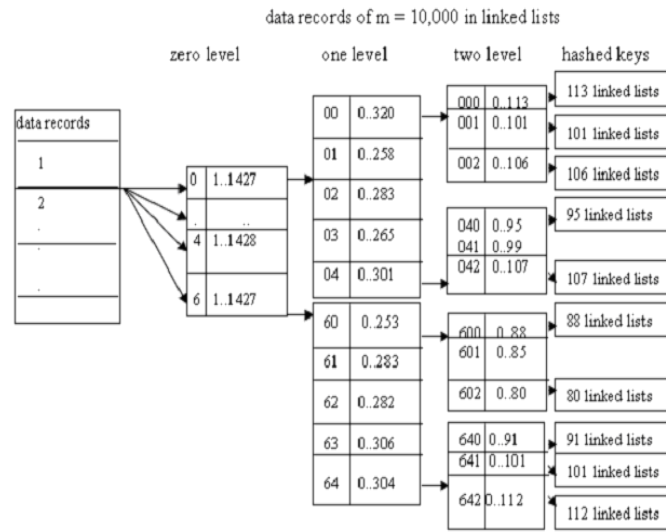

Figure 6: Join operation on multilevel hash tables O (1)